\documentclass[sigconf,screen]{acmart}

\AtBeginDocument{%
  \providecommand\BibTeX{{%
    \normalfont B\kern-0.5em{\scshape i\kern-0.25em b}\kern-0.8em\TeX}}}

\newcommand{\fig}[1]{Figure~\ref{fig:#1}}

\newcommand{\tion}[1]{\S\ref{sec:#1}}
\usepackage{wrapfig}
\usepackage{balance}
\usepackage{multirow}
\usepackage{blindtext, graphicx}
\usepackage{boldline}
\usepackage{subfig} 
\usepackage{booktabs}
\usepackage{ragged2e}
\usepackage{dblfloatfix} 
\usepackage[para,online,flushleft]{threeparttable}
\usepackage{lipsum}

\usepackage[tikz]{bclogo}
\newenvironment{RQ}[1]%
{\noindent\begin{minipage}[c]{0.98\linewidth}%
\begin{bclogo}[couleur=gray!20,%
                arrondi=0.1,%
                logo=\bctrombone,%
                ombre=true]{~#1}}%
{\end{bclogo}\end{minipage}}

\newcommand{\bi}{\begin{itemize}}
\newcommand{\ei}{\end{itemize}}

\usepackage{enumitem}
\setlist[itemize]{leftmargin=*}
\setlist[enumerate]{leftmargin=*}
\usepackage{makecell}
\usepackage[linesnumbered,ruled,vlined]{algorithm2e}
\setlist{nolistsep} 

\SetKwProg{Fn}{Function}{}{}

\setlist[1]{itemsep=0pt}

\usepackage{amsmath}

\usepackage{times}
\usepackage{colortbl}
\usepackage{tikz}
\def\checkmark{\tikz\fill[scale=0.4](0,.35) -- (.25,0) -- (1,.7) -- (.25,.15) -- cycle;}

\makeatletter
\let\th@plain\relax
\makeatother

\definecolor{Gray}{rgb}{0.88,1,1}
\definecolor{Gray}{gray}{0.85}
\definecolor{lightgray}{gray}{0.8}
\usepackage[framed]{ntheorem}
\usepackage{framed}
\usepackage{tikz}
\usetikzlibrary{shadows}
\theoremclass{Lesson}
\theoremstyle{break}

\renewcommand{\textrightarrow}{$\rightarrow$}

\newcommand{\quart}[4]{\begin{picture}(80,4)%1
    {\color{black}\put(#3,2){\circle*{4}}\put(#1,2){\line(1,0){#2}}}\end{picture}}

\def\checkmark{\tikz\fill[scale=0.4](0,.35) -- (.25,0) -- (1,.7) -- (.25,.15) -- cycle;}

\newcommand{\IT}{TERMINATOR}% need a name

\begin{document}
% Title portion 

\copyrightyear{2019} 
\acmYear{2019} 
\setcopyright{acmcopyright}
\acmConference[ESEC/FSE '19]{Proceedings of the 27th ACM Joint European Software Engineering Conference and Symposium on the Foundations of Software Engineering}{August 26--30, 2019}{Tallinn, Estonia}
\acmBooktitle{Proceedings of the 27th ACM Joint European Software Engineering Conference and Symposium on the Foundations of Software Engineering (ESEC/FSE '19), August 26--30, 2019, Tallinn, Estonia}
\acmPrice{15.00}
\acmDOI{10.1145/3338906.3340448}
\acmISBN{978-1-4503-5572-8/19/08}

\title{{\IT}: Better Automated UI Test Case Prioritization }  

\author{Zhe Yu}
\orcid{0000-0002-6841-1725}
\email{zyu9,ffahid,gerother@ncsu.edu, timm@ieee.org}
\author{Fahmid Fahid}
\author{Tim Menzies}
\author{Gregg Rothermel}
\affiliation{%
  \institution{North Carolina State University}
  \postcode{27695}
  \city{Raleigh}
  \state{NC}
  \country{USA}
}

\author{Kyle Patrick}
\email{kyle.patrick@lexisnexis.com}
\author{Snehit Cherian}
\email{snehit.cherian@lexisnexis.com}
\affiliation{%
  \institution{LexisNexis Legal \& Professional}
  \postcode{27606}
  \city{Raleigh}
  \state{NC}
  \country{USA}
}

% \author{Zhe Yu}
% \email{zyu9@ncsu.edu}
% \affiliation{%
%   \institution{North Carolina State University}
%   \postcode{27695}

% \author{Fahmid M. Fahid}
% \email{ffahid@ncsu.edu}
% \affiliation{%
%   \institution{North Carolina State University}
%   \postcode{27695}

% \author{Tim Menzies}
% \email{timm@ieee.org}
% \affiliation{%
%   \institution{North Carolina State University}
%   \postcode{27695}

% \author{Gregg Rothermel}
% \email{gerother@ncsu.edu}
% \affiliation{%
%   \institution{North Carolina State University}
%   \postcode{27695}

% \author{Kyle Patrick}
% \email{kyle.patrick@lexisnexis.com}
% \affiliation{%
%   \institution{LexisNexis Legal \& Professional}
%   \streetaddress{1801 Varsity Dr.}
%   \postcode{27606}

% \author{Snehit Cherian}
% \email{snehit.cherian@lexisnexis.com}
% \affiliation{%
%   \institution{LexisNexis Legal \& Professional}
%   \streetaddress{1801 Varsity Dr.}
%   \postcode{27606}

\renewcommand{\shortauthors}{Yu et al.}

\begin{abstract}
Automated UI testing is an important component of the continuous integration process of software development. A modern web-based UI is an amalgam of reports from dozens of microservices written by multiple teams. Queries on a page that opens up another will fail if any of that page's microservices fails. As a result, the overall cost for automated UI testing is high since the UI elements cannot be tested in isolation. For example, the entire automated UI testing suite at LexisNexis takes around 30 hours (3-5 hours on the cloud) to execute, which slows down the continuous integration process.

To mitigate this problem and give developers faster feedback on their code, test case prioritization techniques are used to reorder the automated UI test cases so that more failures can be detected earlier. Given that much of the automated UI testing is ``black box'' in nature, very little information (only the test case descriptions and testing results) can be utilized to prioritize these automated UI test cases. Hence, this paper evaluates 17 ``black box'' test case prioritization approaches that do not rely on source code information. Among these, we propose a novel TCP approach, that dynamically re-prioritizes the test cases when new failures are detected, by applying and adapting a state of the art framework from the total recall problem. Experimental results on LexisNexis automated UI testing data show that our new approach (which we call {\IT}), outperformed prior state of the art approaches in terms of failure detection rates with negligible CPU overhead.
\end{abstract}




\maketitle


\section{Introduction}
\label{sec:introduction}

% What is automated UI testing?, Why is it important?
Complex web applications must be  validated using a variety of methods.  
Early in the lifecycle, unit and integration tests can be used to validate many aspects of an application. However, these isolated tests cannot realistically simulate the behavior of a modern web page that combines together many services. 
For this reason, user interface (UI) testing are applied and automated. Compared to unit tests, automated UI tests are more expensive to write and maintain. Also, they usually have a higher execution time.  
 For example, the automated UI test suite at LexisNexis
  takes around 30 hours (3-5 hours on the cloud) to execute. Hence:
\bi
\item Testing has become the key component in continuous integration. If a   system can only be rebuilt and tested (say) twice in an eight-hour work day, that also determines how often developers can ship new (or fixed) features to clients.
\item
Developers are less agile when testing forces them to wait for hours to get feedback on their latest changes.  

\begin{figure*}
\vspace{5mm}
    \centering
    \includegraphics[width=\linewidth]{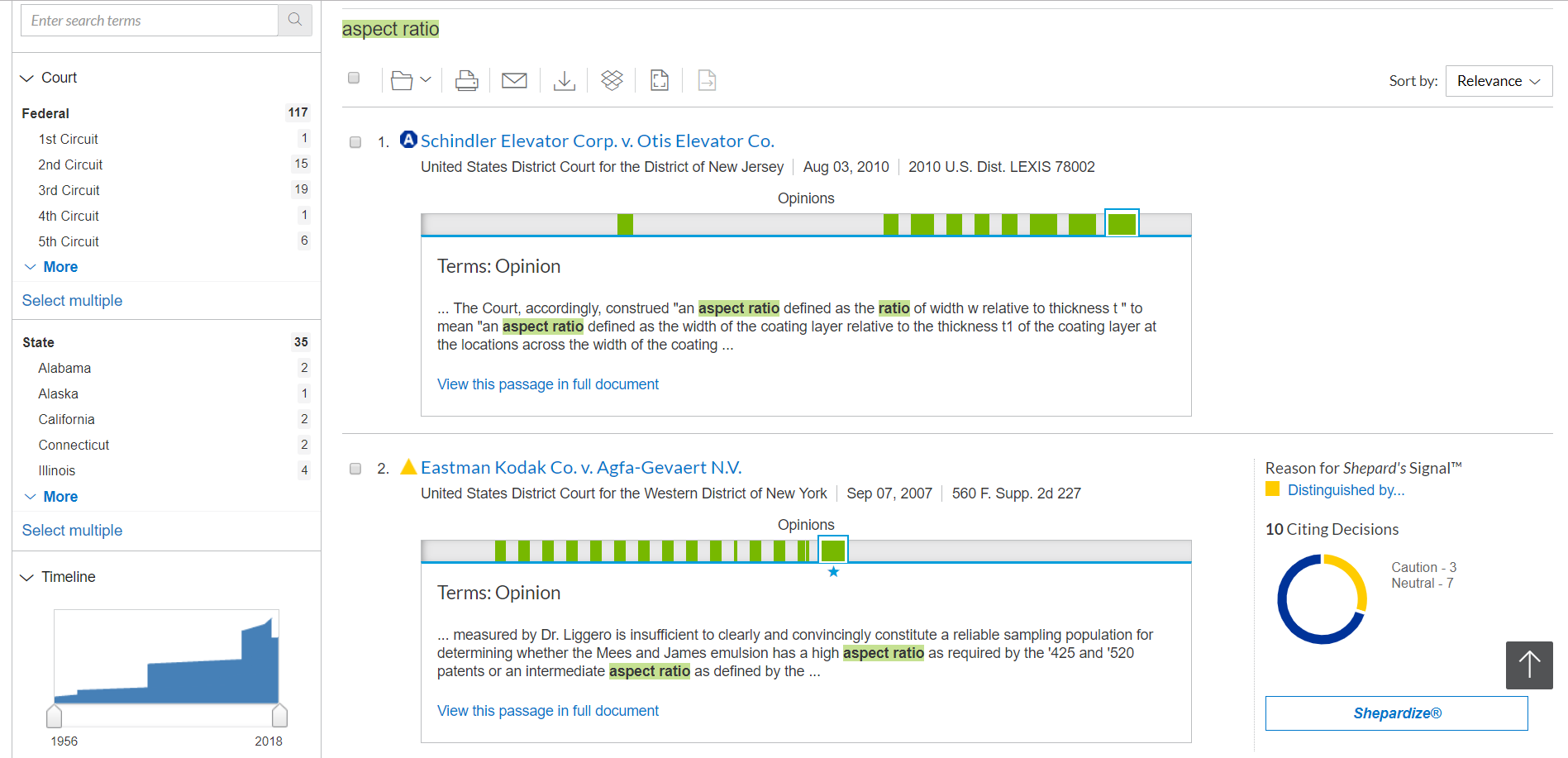}
    \caption{A modern web page is an amalgam of many microservices.
    This figure  shows a real example from LexisNexis where a single click  invokes multiple components of the software.
    By our estimates, the code on this page comes from seven different teams which, in turn, use ``under-the-hood'' components from dozens of other teams.
     The   failure of any component will lead to a test failure. That
     failure may be opaque to the author of the test since it  may result
     from the failure of components  written by many other teams. 
    As a result, when a team is prioritizing their own tests,
    all they can reason about is  the test description information  (e.g see \fig{example}(b))  and prior results from their own testing  results (time to run the test;  whether or not that prior test failed).    }
    \label{fig:LN}
\end{figure*}

\ei 
Worse still,  since automated UI tests are expressed in terms of actions taken by a browser user agent,   failures do not  have  a straightforward relationship to the underlying application code or architecture.  As shown in \fig{LN}, a modern web-based UI is an amalgam of reports from dozens of microservices written by multiple teams. If a query on one webpage opens up another, then that query interacts with code from multiple teams.  
The code that generated  \fig{LN}  comes from seven different teams which, in turn, use ``under-the-hood'' components from dozens of other teams.
Consequently, much of the testing is ``black box'' in nature since most of the code
associated with subpages will be opaque to the query author. Hence, it is hard
to determine what part of the code is being tested by which test case.

\begin{figure*}
    \vspace{5mm}
    \centering
    \includegraphics[width=\linewidth]{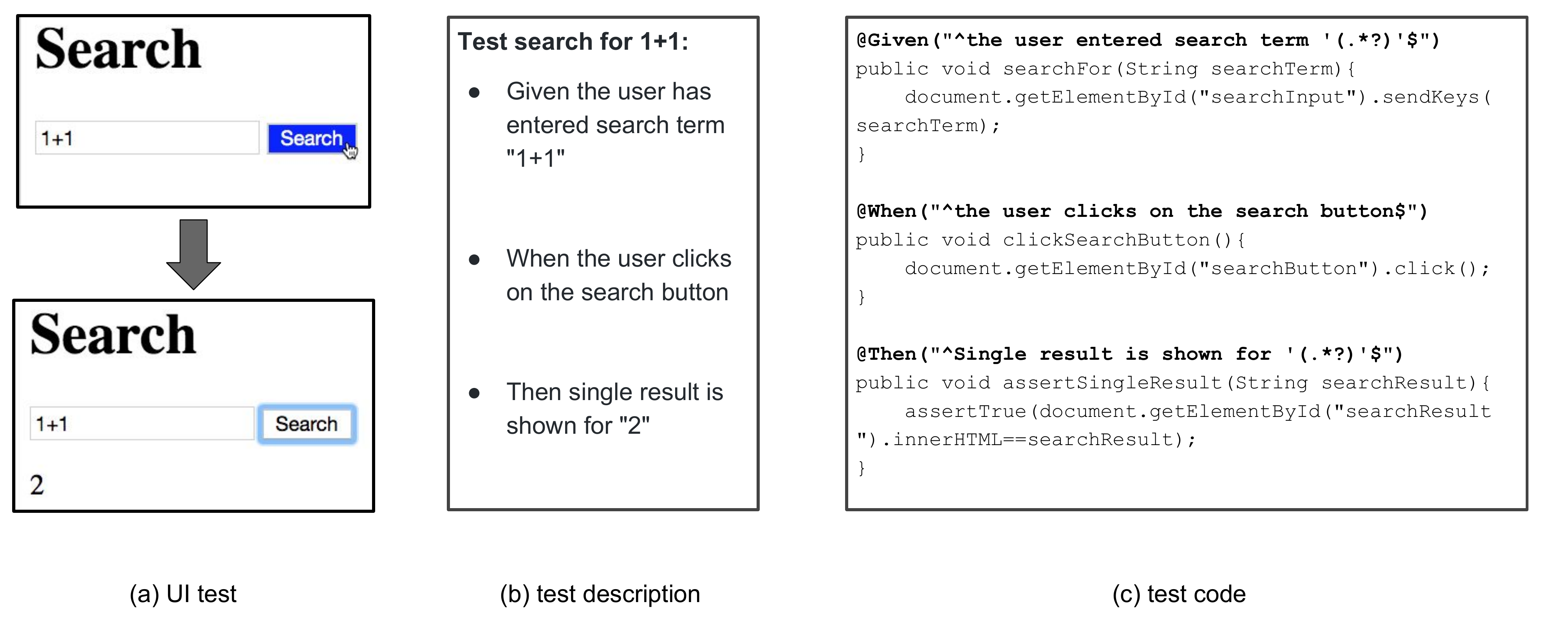}
    \caption{To test a page shown at the left~(a), programmers write a test description~(b) which is converted to test code~(c).}
    \label{fig:example}
\end{figure*}
 
% error messages. Meanwhile, the triage also consumes hours of human and machine resources in order to figure out what part of the code is responsible for the test failures.

% how to fix the problem?
One possible solution to this problem is test case prioritization (TCP). TCP is a topic widely studied in software testing, where test cases are prioritized so that test cases revealing faults are executed earlier, allowing software developers to identify and fix problems earlier~\cite{4,5,6,7,8}.
A problem with many of the existing TCP techniques is that they 
rely on coverage information extracted from the source code, such as what branches/statements/functions of the source code are covered by which test cases~\cite{78,80,89,91,92}. Unfortunately, such techniques are not applicable to automated UI testing since the coverage information is not available---as discussed above, the relationship between an automated UI test case and the source code being tested is blurred.

This paper explores TCP techniques for higher failure detection rates utilizing only ``black box'' information; i.e. test descriptions and results from their own prior (and current) test runs (time to run the test, whether or not that test failed). Our central insight  is that this kind of test case prioritization problem belongs to a class of information retrieval problems called the {\em total recall} problem~\cite{grossman2016trec} (defined in \tion{total recall}). 
To test this idea, in this work we apply total recall methods to ``read'' the 
description and history results of the test cases to build a classifier that predicts which tests might fail sooner.  Specifically, we incrementally update a support vector machine (SVM) model with the results of executed tests, then reflect on the SVM model to decide which test to run next. Formally speaking, this is an {\em active learning}~\cite{settles2012active} based framework.
 
We compared our proposed approach, which we call {\IT}, with  12 state of the art TCP algorithms and 2 baseline algorithms on 54 consecutive runs of automated UI testing   (data from three months of testing at LexisNexis) to answer the following two research questions:
\bi
\item
\textbf{RQ1: can the proposed approach {\IT} achieve significantly higher failure detection rates than other TCP algorithms when prioritizing automated UI tests from LexisNexis?} Our results show that {\IT} significantly outperformed other TCP algorithms in terms of failure detection rates. This suggests that techniques from the total recall problem can be successfully applied and adapted to address the automated UI test case prioritization problem.
\item
\textbf{RQ2: what is the computational overhead of {\IT}?} Our results show that {\IT} had 50\% more computational overhead than the simple history-based TCP algorithms because it recursively updates the SVM model and dynamically adjusts the order of the unexecuted tests. However, {\IT}'s overhead was still negligible compared to the runtime of the test cases (0.33\% of the runtime of the test suite). Therefore, we conclude that it is practical to apply {\IT} for prioritizing automated UI tests in LexisNexis.
\ei
The main contributions of this paper are:
\begin{enumerate}
\item
Identification of an under-explored problem: prioritizing automated UI tests. Features of this problem are: (a) the prioritization target is to reach higher failure detection rates and (b) source code  information is unavailable. By our reading of the  literature, most existing TCP approaches utilize source code information and prioritize for fault detection rates. Only 27 of 239 approaches do not rely on source code information, among which we derived 12 approaches that can prioritize automated UI tests.
\item
A dataset for automated UI testing is provided\footnote{https://github.com/ai-se/Data-for-automated-UI-testing-from-LexisNexis} for reproducing and improving this work. 
This is an important contribution since, in the research literature, there are very few examples of real-world industrial test results from large systems.
\item
The lesson learned  that test case prioritization for failure detection rates can be generalized as the total recall problem. 
\item
By applying and adapting the total recall problem, we propose a new TCP algorithm that significantly outperformed other TCP algorithms in terms of failure detection rates.
% \item
% The impact of the proposed algorithm is immediate, LexisNexis blah blah
\item
The proposed algorithm can be easily applied or tested on other TCP problems since it requires minimal information.
\end{enumerate}

The rest of this paper is structured as follows. Background and related work on automated UI testing and test case prioritization are discussed in \tion{background}. The total recall problem is introduced and the proposed new TCP algorithms are presented in \tion{methodology}. The proposed TCP approach {\IT} is compared against other TCP algorithms in \tion{experiments}. Impacts of this work to LexisNexis are reported in \tion{impact}
% . Threats to validity to this work are analyzed in \tion{validity}
while conclusion and future work are provided in \tion{conclusion}.

\section{Background and Related Work}
\label{sec:background}

LexisNexis is a corporation providing computer-assisted legal research (CALR) as well as business research and risk management services~\cite{lnnyt,lnnyt1}. During the 1970s, LexisNexis pioneered the electronic accessibility of legal and journalistic documents~\cite{lnnyt2}. As of 2006, the company had the world's largest electronic database for legal and public-records related information~\cite{lnnyt2}.

LexisNexis provides regulatory, legal, business information and analytics to the legal community.  Legal and research professionals use the Lexis Advance platform to find relevant information more easily and efficiently~\cite{lnlegal}. It helps them prepare legal cases and drive better legal outcomes. The Lexis Advance platform is maintained by a set of automated UI tests. Those automated UI tests simulate user behaviours on the interface of the platform and detect potential failures of the underlying microservices whenever the system is modified and rebuilt.

\subsection{Automated UI testing}
\label{sec:problem}

Automated UI testing is an important component of the continuous integration process of  software development. To automatically test the user interface of a complex system such as the Lexis Advance platform, the test team of LexisNexis designs a large set of automated UI test cases, each of which simulates one action in a specific scenario. As an example, Figure~\ref{fig:example} shows how one automated UI test case is designed for a simple search of "1+1". In this example, the test designer wants to test the search function by (a) verifying that when a user inputs "1+1" and hits the search button, a result of "2" will show up. To automate this UI test, the test designer would first (c) define the test code for a set of scenarios, then (b) write down the automated UI test case with the pre-defined scenarios and expected input and output. In this way, the test designer does not need to know what code will be executed when an automated UI test is executed, and the pre-defined scenarios can be reused for designing other automated UI test cases.

\subsection{Test Case Prioritization}
\label{sec:TCP}

The goal of test case prioritization is to schedule test cases for execution in an order that attempts to increase their effectiveness at meeting some performance goal~\cite{167}.
Here we briefly introduce different types of TCP algorithms in terms of what they prioritize for and what information they use.

\subsubsection{What to prioritize for}
\label{sec:what to prioritize for}
Three different performance goals have been  targeted by TCP algorithms:
\bi
\item
\textbf{Coverage:} Some approaches aim to order test cases so that higher coverage can be achieved earlier. Here the coverage can refer to requirement coverage~\cite{128}, statement coverage~\cite{89}, decision coverage~\cite{89}, block coverage~\cite{89}, branch coverage~\cite{167}, etc.
\item
\textbf{Fault detection rate:} Most approaches prioritize the test cases to achieve higher fault detection rates. An improved rate of fault detection during testing can provide faster feedback on the system under test and let software engineers begin correcting faults earlier than might otherwise be possible~\cite{167}. The most popular performance metric for evaluating early fault detection is average percentage of faults detected (APFD) proposed by Rothermel et al.~\cite{167} in 2001. APFD computes the area under the curve (AUC) of the the gain in the percentage of detected faults as follows:
\begin{equation}\label{eq:APFD}
APFD = 1-\frac{TF1+TF2+\cdots+TFm}{nm}+\frac{1}{2n}
\end{equation}
where $TFi$ is the first test case that reveals fault $i$, $m$ is the total number of faults revealed by the test suite, and $n$ is the total number of test cases in the test suite. Ranging from 0\% to 100\%, higher APFD values mean better ordering of test cases in terms of early fault detection. 

In APFD, each test case is considered to be of the same cost and the same severity. However in practice, some test cases can take much longer to run than others while some faults can cause more damage than others if not detected. To this end, Elbaum et al.~\cite{145} proposed the metric of average percentage of faults detected with cost (APFDc), which takes into consideration the cost related to the
resources required to execute and validate each test case and the severity of each fault. Let $T$ be a test suite containing $n$ test cases with costs $t1, t2, \dots, tn$.  Let $F$ be a set of $m$ faults revealed by the test suite and let $f1, f2, \dots,fm$ be the severities of those faults. Let $TFi$ be the first test case that reveals fault $i$. APFDc is calculated as follows:
\begin{equation}\label{eq:APFDc}
APFDc = \frac{\sum_{i=1}^m(fi\times(\sum_{i=TFi}^nti-0.5t_{TFi}))}{\sum_{i=1}^nti\times \sum_{i=1}^{m}fi}
\end{equation}
In order to measure APFD or APFDc, researchers use manually seeded faults~\cite{118}, mutation faults~\cite{41,42} or real fault detection records~\cite{192,197}. Source code information is required to manually seed faults or to generate mutation faults while failure to fault mapping information (i.e. which failures reveal which faults) is required for real fault detection records.
\item
\textbf{Failure detection rate:} Some studies prioritize test cases to achieve higher failure detection rates. This is a compromise prioritization target when failure to fault mapping information is not available and is often seen when real world data is applied to evaluate TCP performance. Similar to fault detection rate, Liang et al.~\cite{google2} apply APFDc to measure failure detection rates. Here it becomes average percentage of failures detected with the same equation as \eqref{eq:APFDc}, except that $TFi$ represents $i$th failed test case. 
\ei

\subsubsection{What information is used}
\label{sec:what information is used}
Different TCP algorithms utilize different information to decide the execution order of test cases:
\bi
\item
\textbf{Coverage information:} 
Most test case prioritization techniques rely on coverage information extracted from source code~\cite{167}, changes in source code~\cite{52}, requirements~\cite{111}, previously detected faults~\cite{33}, etc. Usually, coverage information requires access to the source code and takes time to compute.
\item
\textbf{History information:} 
Some algorithms learn the fault/failure exposing potential of each test case from its fault/failure detection history to prioritize test cases for current execution. Then, the test cases can be prioritized in descending order of their fault/failure exposing potential.
\item
\textbf{Cost information:} 
Some TCP algorithms take into consideration the cost of each test case to prioritize the test cases, e.g. execute all test cases in ascending order of their runtime estimated from the execution history~\cite{cost-based}. 
\item
\textbf{Test description information:} 
Standard natural language processing techniques (stopword removal, stemming, bag-of-word, etc.) can be applied to extract features from test description information as shown in Figure~\ref{fig:example}(b). Some algorithms train a predictive model with independent variables from test description features and dependent variables extracted from failure/fault history. The test cases are then prioritized based on the predicted probability of failure/fault detection~\cite{39,59}. 
\item
\textbf{Feedback information:}
Some algorithms iteratively utilize execution results (pass/fail) in current test runs to reorder the rest of the unexecuted test cases in the test suite~\cite{183,103}. The priorities of  unexecuted test cases are adjusted dynamically based on two heuristics: (1) if one test case fails, increase the priority of unexecuted test cases that are similar/related to the failed one; (2) if one test case passes, decrease the priority of unexecuted test cases that are similar/related to the passed one. 
\ei

\subsection{Prioritizing automated UI test cases}
\label{sec:TCP for UI}

There are two characteristics of automated UI test case prioritization: (1) its prioritization target is to reach higher failure detection rate, and (b) source code  information is unavailable. 
To analyze what existing TCP algorithms can be applied to prioritize automated UI test cases, we conducted a systematic literature review~\cite{YuThesis} and ended up identifying 239 TCP papers. Surprisingly, we found that there are few papers~\cite{183,212,232,59} (4 out of 239) exploring similar scenarios as automated UI test case prioritization---prioritizing for failure detection rates without source code information. This suggests that although lots of work has been done in the TCP arena, the scenario where source code information and failure to faults mapping are not available is under-explored and many of the existing TCP algorithms cannot be applied to prioritizing automated UI tests.

\section{Methodology}
\label{sec:methodology}

The central idea  of this paper is that  prioritizing automated UI tests can be generalized as the total recall problem~\cite{grossman2016trec} and techniques addressing the total recall problem can be applied and adapted to better prioritize the automated UI tests. 

\subsection{Total Recall}
\label{sec:total recall}

The aim of {\em total recall} is to optimize the cost for achieving very high recall (ideally, very close to 100\%) with a human assessor in the loop~\cite{grossman2016trec}. More specifically, the total recall problem can be described as follows:

\begin{RQ}{The Total Recall Problem:}
  Given  candidates $E$ with a small positive fraction $R \subset E$ , each  $x \in E$ can be inspected to  label it positive ($x\in R$) or negative ($x \not\in R$) at a cost. Starting with the labels $L = \emptyset$, the task is to inspect and label as few candidates as possible ($\min |L|$) while achieving very high recall ($\max |L\cap R|/|R|$).
\end{RQ}

State of the art solutions~\cite{Yu2018,Yu2019} for the total recall problem use an active learning based framework. The key idea behind active learning
\begin{wrapfigure}{r}{1.6in}
\includegraphics[width=1.55in]{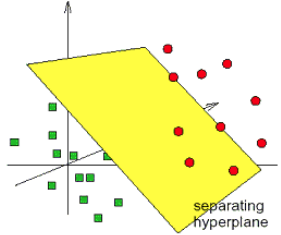}
\caption{Separating failing (red circles) from  passing (green squares) test cases.}\label{fig:svm}
\end{wrapfigure}
is that a machine learning algorithm can train  faster (i.e. using less  data) if it is allowed to choose the data from which it learns~\cite{settles2012active}. 
The experience in explored total recall problems is that  such {\em active learners} outperform supervised and semi-supervised learners and can significantly reduce the effort required to achieve high
recall~\cite{Cormack2017Navigating,Cormack2016Engineering,cormack2016scalability,cormack2015autonomy,cormack2014evaluation,grossman2013,wallace2010semi,wallace2010active,wallace2011should,wallace2012class,wallace2013active,Yu2018,Yu2019}.
To understand active learning, consider the decision plane between the failed and   passed test cases of \fig{svm}. Suppose we want to find more failing test cases and we had access to the  \fig{svm} model. One tactic would be to execute test cases that fall into the  region of red circles in this figure, as far as possible from the green squares (this tactic is called {\em certainty sampling}). Another tactic would be to verify the position of the boundary; i.e. execute test cases that are closest to the boundary (this tactic is called {\em uncertainty sampling}).

% \begin{figure}[!b]
%     \centering
%     \includegraphics[width=\linewidth]{General_FASTREAD.pdf}
%     \caption{Block diagram of the active learning based framework for total recall problems}
%     \label{fig:fastread}
% \end{figure}

\subsection{{\IT}}
\label{sec:method}

The automated UI test case prioritization problem can be generalized to the total recall problem as follows:
\bi
\item 
$E$: the test suite to be prioritized.
\item
$R$: the set of test cases that will fail if executed.
\item
$L$: the set of test cases already executed in the current run.
\item
$L_R = L\cap R$: the set of failed test cases in the current run.
\ei
Given the information available in automated UI testing, we extract three types of features:
\bi
\item
\textbf{Text feature}: the same text mining feature extraction used in the total recall approaches~\cite{Yu2018,Yu2019} is applied to extract text features from the test case descriptions. Specifically, we:
\begin{enumerate}
\item
Tokenized the test case descriptions without stop/control word removal or stemming.
\item
Built a term frequency matrix.
\item
Normalized each row (feature vector for each test case) with their L2-norm\footnote{
L2-norm for a vector $x$ is $\sqrt{x^{T}x}$, where $T$ denotes ``transpose''}.
\end{enumerate}
\item
\textbf{History feature}: same as history-based algorithms, the testing result from previous runs (failed, passed, skipped) are vectorized as feature.
\item
\textbf{Hybrid feature}: concatenation of text and history features.
\ei
Using the foregoing types of features, the proposed framework is described in Algorithm~\ref{alg:alg} with engineering choices of $N_1,N_2$. $N_1$ is the batch size of the process. In this paper, we chose $N_1=10$ to simulate the scenario where test cases are executed in parallel on 10 computation nodes on the cloud. $N_2$ represents the threshold above which certainty sampling is applied instead of uncertainty sampling. In this paper, we chose $N_2=30$ as suggested by previous works on total recall~\cite{Yu2019}.

\begin{algorithm}[!tbh]
\scriptsize
\SetKwInOut{Input}{Input}
\SetKwInOut{Output}{Output}
\SetKwInOut{Parameter}{Parameter}
\SetKwRepeat{Do}{do}{while}
\Input{$E$, the test suite\\$R$, test cases that will fail\\$N_1$, batch size\\$N_2$, threshold of query strategy}
\Output{$L$, list of executed test cases\\$L_R$, list of failed test cases}
\BlankLine

$L\leftarrow \emptyset$\; $L_R\leftarrow \emptyset$\; 

\BlankLine
\tcp{Keep reviewing until all the test cases are executed}
\While{$|L| <  |E|$}{
    \tcp{Start training or not}
    \eIf{$|L_R| \geq 1$}{
        \tcp{Presumptive non-relevant examples}
        $L_{\text{pre}}\leftarrow \text{Presume}(L,E\setminus L)$\;
        $CL\leftarrow \text{Train}(L_{\text{pre}})$\;
        \tcp{Query next}
        $X\leftarrow \text{Query}(CL,E\setminus L,L_R)$\;
    }{
        \tcp{Random Sampling}
        $X\leftarrow \text{Random}(E\setminus L)$\;
    }
    \tcp{Execute the selected test cases}
    \ForEach{$x \in X$}{
      $L_R,L\leftarrow \text{Execute}(x,R,L_R,L)$\;
    }
}
\Return{$L,L_R$}\;
\BlankLine
\Fn{Presume ($L,E\setminus L$)}{
    \tcp{Randomly sample $|L|$ points from $E\setminus L$, presume those to be passed test cases}
    \Return $L \cup \text{Random}(E\setminus L,|L|)$\;
}
\BlankLine
\Fn{Train ($L_{\text{pre}}$)}{
    \tcp{Train linear SVM with Weighting}
    $CL\leftarrow \text{SVM}(L_{\text{pre}},\text{kernel=linear},\text{class}\_\text{weight=balanced})$\;
    \If{$L_R\ge N_2$}{
        \tcp{Aggressive undersampling}
        $L_I\leftarrow L_{\text{pre}}\setminus L_R$\;
        $\text{tmp}\leftarrow L_I[\text{argsort}(CL.\text{decision}\_\text{function}(L_I))[:|L_R|]]$\;
        $CL\leftarrow \text{SVM}(L_R \cup \text{tmp, kernel=linear})$\;
    }
    \Return{$CL$}\;
}
\BlankLine
\Fn{Query ($CL,E\setminus L,L_R$)}{
    \eIf{$L_R\ge N_2$}{
        \tcp{Certainty Sampling (highest predicted probability of failing)}
        $X\leftarrow \text{argsort}(CL.\text{decision}\_\text{function}(E\setminus L))[::-1][:N_1]$\;
    }{
        \tcp{Uncertainty Sampling}
        $X\leftarrow \text{argsort}(\text{abs}(CL.\text{decision}\_\text{function}(E\setminus L)))[:N_1]$\;
    }
    \Return{$X$}\;
}
\BlankLine
\Fn{Execute ($x,R,L_R,L$)}{
    \tcp{Append selected test case to the list of executed ones}
    $L\leftarrow [L, x]$\;
    \tcp{If the selected test case fails, append it to the list of failed ones}
    \If{$x\in R$}{
        $L_R\leftarrow [L_R, x]$\;
    }
    \Return{$L_R$, $L$}\;
}
\caption{Pseudo Code for {\IT}}\label{alg:alg}
\end{algorithm}

\section{Empirical Study}
\label{sec:experiments}

We conducted an empirical study to answer the following research questions:
\bi
\item
\textbf{RQ1: can {\IT} achieve significantly higher failure detection rates than other TCP algorithms when prioritizing automated UI tests from LexisNexis?} Given that the goal of automated UI test case prioritization is to detect failures faster, {\IT} will only be considered useful it can achieve significantly higher failure detection rates than other TCP algorithms.
\item
\textbf{RQ2: what is the computational overhead of {\IT}?} It is also important for a TCP algorithm to have low computational overhead, e.g., a TCP algorithm will not be considered useful if it can sort test cases in an effective order but requires even longer time than the total runtime of the test suite  to compute that order.
\ei

\subsection{Dataset}

To conduct this study, we collected data for 54 consecutive runs of 2661 automated UI test cases in LexisNexis from  September 27th to November 15th in 2018. The collected data contains the test description written using Gherkin syntax\footnote{https://cucumber.io/docs/gherkin/} (as shown in Figure~\ref{fig:example} (b)), test duration, outcome (passed/failed), and error message associated with each test case. To remove the effect of environmental errors, failures with "time out" messages are labeled with "Timeout" and are not considered failures. This dataset is available online\footnote{https://github.com/ai-se/Data-for-automated-UI-testing-from-LexisNexis}. For confidentiality, only the featurized vectors are provided for test descriptions while error messages are omitted.

The collected dataset is used to simulate the performance of different TCP algorithms. We simulate Run 6 to Run 54 by using different prioritization approaches and compare their performance\footnote{Run 1 to Run 5 are not simulated since there are not enough execution history information available for those runs.}. When prioritizing for Run n, the execution results from Run 1 to Run n-1 are available as information for the prioritization algorithm. 

Table~\ref{tab:example} shows a small example of the dataset with 4 test cases and 4 consecutive runs. To prioritize for test cases in Run 4, all testing results from Run 1 to Run 3 and  the description of each test case can be utilized.

\begin{table}
\caption{A simple example of automated UI test case prioritization, where P, F, and S indicate passed, failed, and skipped testing results. For each test session, the runtime is t1<t2<t3<t4.}
\label{tab:example}
\begin{center}
% \footnotesize
\setlength\tabcolsep{8pt}
\begin{tabular}{ll|llll}
&& \multicolumn{4}{c}{Test Session}\\\hline
Test	& Description 	&1	&2	&3	&4         \\\toprule
t1&	Test Check Box in  page A  	&P	&F	&S	&F \\
t2&	Test Radio Button in  page A	&F&	F&	P&	F \\
t3&	Test Check Box in  page B	&P	&P	&F	&P \\
t4&	Test Radio Button in  page B	&F	&P	&F	&F \\\bottomrule
\end{tabular}
\end{center}
\end{table}

\begin{table*}
\caption{Test Case Prioritization Algorithms and Information They Utilize}
\label{tab:algorithms}
\begin{center}
\begin{threeparttable}
\footnotesize
\setlength\tabcolsep{4pt}
\begin{tabular}{c|p{10.5cm}|ccc|}
\multicolumn{2}{c|}{~}                & \multicolumn{3}{c|}{\textbf{Utilized Information}}\\\cline{3-5}
\textbf{ID}     & \textbf{Algorithm Description}       & Execution history     & Test case description       & Feedback       \\ \hline
A1 & Execute test cases in random order. & & &  \\ 
A2 & Execute test cases in optimal order. & & &  \\ \hline
B1 & Execute test cases in ascending order of time since last failure. & \checkmark & &  \\ 
B2 & Execute test cases in descending order of number of times failed/number of times executed. & \checkmark & &  \\ 
B3 & Execute test cases in descending order of exponential decay metrics as in \eqref{eq:Q-learning}. & \checkmark & &  \\ 
B4 & Execute test cases in descending order of ROCKET metrics as in \eqref{eq:ROCKET}. & \checkmark & &  \\  
B5 & Execute test cases in descending order of the Mahalanobis distance of each test case to the origin (0,0) when considering two metrics---time since last execution and failure rate. & \checkmark & &  \\\hline
C1 & Execute test cases in ascending order of the estimated test case runtime. & \checkmark & &  \\ \hline
D1 & Supervised learning with Simple History (SH). & \checkmark & \checkmark &  \\ 
D2 & Supervised learning with All History (AH). & \checkmark & \checkmark &  \\ 
D3 & Supervised learning with Weighted History (WH). & \checkmark & \checkmark &  \\ \hline
E1 & Dynamic test case prioritization with co-failure information. & \checkmark &  & \checkmark \\ 
E2 & Dynamic test case prioritization with flipping history. & \checkmark &  & \checkmark \\ 
E3 & Dynamic test case prioritization with rules mined from failure history. & \checkmark &  & \checkmark \\  \hline
F1 & {\IT} with text feature. &  & \checkmark & \checkmark \\ 
F2 & {\IT} with history feature. & \checkmark &  & \checkmark \\ 
F3 & {\IT} with hybrid feature. & \checkmark & \checkmark & \checkmark \\ \hline
\end{tabular}
% \begin{tablenotes}\small
% This table presents the test case prioritization algorithms tested in this paper. Group A are baselines with A1 being a lower bound of any effective prioritization algorithm and A2 being the upper bound.  Group B are history-based and cost-based algorithm. Croup C is cost-based algorithm. Group D are test case-based algorithms. Group E are feedback-based algorithm. Algorithms from Group B to E are existing state of the art algorithms. Group F are the proposed algorithms in this paper.
% \end{tablenotes}
\end{threeparttable}
\end{center}

\end{table*}

\subsection{Independent Variables}
\label{sec:treatments}

Three variants  of the proposed active learning based TCP framework {\IT} are compared against 12 existing ``black box'' TCP techniques and 2 baselines. As shown in Table~\ref{tab:algorithms}, in total 17 algorithms are tested in which Group A are baseline algorithms, Groups B, C, D, and E are the existing ``black box'' TCP techniques, and Group F are three variants  of {\IT} that differs from what types of features  they use. The 12 TCP techniques are chosen because they do not rely on source code information. During the simulation, three types of information can potentially be applied by these techniques to prioritize the test cases in Run n:
\bi
\item
\textbf{Execution history:} execution results of each test case from Run 1 to Run n-1. For example the execution history information for t1 in Run 4 would be \{P, F, S\} in Table~\ref{tab:example}.
\item
\textbf{Test case description:} the natural language description of the automated UI test case. For example the test case description information for t1 would be ``Test Check Box in page A'' in Table~\ref{tab:example}.
\item
\textbf{Feedback:} execution results of prior test cases in Run n. For example in Table~\ref{tab:example}, in Run 4, when t1 has been executed and other tests have not, the feedback information would be ``t1 failed''. 
\ei
Note that some of the state of the art algorithms are not reproduced exactly as presented in the original paper, rather, they are partially implemented because of lack of information or differences in the workflow of automated UI testing.

\subsubsection{Group A: baselines}

Group A shows baseline algorithms for comparison, in which 
\bi
\item
\textbf{A1} is the simplest TCP algorithm utilizing no information. It executes all the test cases in a random order. A TCP algorithm has no value if it cannot perform better than A1. In the example of Table~\ref{tab:example}, A1 could place tests in an order such as \{t1, t3, t2, t4\} for Run 4. 
\item
\textbf{A2} executes the test cases in the optimal order. It represents the upper bound of the TCP performance any algorithm can ever achieve. In the example of Table~\ref{tab:example}, A2 will execute the tests in the order \{t1, t2, t4, t3\} for Run 4 to maximize the failure detection rate. A2 is not applicable in practice because it requires knowledge of which test cases will fail to determine the optimal order, but it can be used in simulations to show how much room there is to improve from the TCP algorithms.
\ei

\subsubsection{Group B: history-based algorithms}

Group B algorithms use metrics extracted from the execution history information, as described in \tion{what information is used} to order the test cases before each run:
\bi
\item
\textbf{B1: }time since last failure. Many algorithms assign higher priority to test cases with more recent failures~\cite{38,google,79}. To apply this metric to our dataset, it becomes the number of consecutive non-failures before Run n. For example, B1 metrics for Run 4 in Table~\ref{tab:example} are \{1, 1, 0, 0\} for t1 to t4. Therefore, B1 will sort the test cases in the order \{t3, t4, t1, t2\} before Run 4.
\item
\textbf{B2: }failure rate (number of times failed/number of times executed). Fazlalizadeh et al.~\cite{7}, Aman et al.~\cite{232}, and Tsai et al.~\cite{212} apply this metric to determine the order of test cases (higher priority for higher value of B2 metrics). For example, B2 metrics for Run 4 in Table~\ref{tab:example} are \{1/2, 2/3, 1/3, 2/3\} for t1 to t4. Therefore, B2 will sort the test cases in the order \{t2, t4, t1, t3\} before Run 4.
\item
\textbf{B3: }Exponential decay metrics~\cite{97}
\begin{equation}\label{eq:Q-learning}
\begin{aligned}
&P_0 = h_1 \\
&P_i = \alpha h_{i} + (1-\alpha) P_{i-1}, 0\le \alpha \le 1, i \ge 1
\end{aligned}
\end{equation}
where $h_i=0$ if the test case passed or was skipped in Run i and $h_i=1$ if the test case failed in Run i. $\alpha$ is the learning rate. In our experiments, $\alpha=0.9$ achieves better performance than other values ranging from $0.1$ to $1$. Higher priority is assigned for higher B3 metrics. For example, B3 metrics for Run 4 in Table~\ref{tab:example} are \{0.09, 0.1, 0.9, 0.99\} for t1 to t4. Therefore, B3 will sort the test cases in the order \{t4, t3, t2, t1\} before Run 4.
\item
\textbf{B4: }ROCKET metrics~\cite{182}
\begin{equation}\label{eq:ROCKET}
\begin{aligned}
&P_i = \sum_{j=1}^{i-1}{\omega_{i-j}h_{j}} \\
&\omega_{k} = \left\{  
             \begin{aligned}
             0.7, & \quad \text{if } k=1   \\  
             0.2, & \quad \text{if } k=2   \\  
             0.1, & \quad \text{if } k\ge 3   
             \end{aligned}  
            \right.
\end{aligned}
\end{equation}
where $h_j=0$ if the test case passed or was skipped in Run j and $h_j=1$ if the test case failed in Run j. Higher priority is assigned for higher B4 metrics. For example, B4 metrics for Run 4 in Table~\ref{tab:example} are \{0.2, 0.3, 0.7, 0.8\} for t1 to t4. Therefore, B4 will sort the test cases in the order \{t4, t3, t2, t1\} before Run 4.
\item
\textbf{B5: }time since last execution. Some algorithms assign higher priority to test cases that have not been executed for a long time~\cite{232,google}. To apply this metric to our dataset, we utilized the number of consecutive skips before Run n. For example, using this metric, the value for Run 4 in Table~\ref{tab:example} are \{1, 0, 0, 0\} for t1 to t4. Given that there are only a few skipped test cases in our dataset, this metric is never applied alone. Therefore, we followed the work of Aman et al.~\cite{232} using time since last execution and failure rate (B5) to prioritize the test cases by the Mahalanobis distance of the two metrics to the origin:
\begin{equation}\label{eq:B5}
\begin{aligned}
d_M(\mathbf{x},(0,0)) = \mathbf{x}^{T}S^{-1}\mathbf{x}
\end{aligned}
\end{equation}
where $S$ is the variance-covariance matrix of all the data $\mathbf{x}$ and $S^{-1}$ is its inverse. For example, B5 metrics for t1 to t4 in Run 4 are \{(1, 1/2), (0, 2/3), (0, 1/3), (0, 2/3)\}. Their variance-covariance matrix and its inverse are:
\begin{equation*}
S = \begin{bmatrix} 
0.250 & -0.014  \\ 
-0.014 & 0.025  
\end{bmatrix}, \text{ and }S^{-1} = \begin{bmatrix} 
4.124 & 2.25  \\ 
2.25 & 40.5
\end{bmatrix}
\end{equation*}
Their Mahalanobis distances to the origin are \{17.625, 18, 4.5, 18\}. Therefore, B5 will sort the test cases in the order \{t2, t4, t1, t3\} before Run 4.
\ei

\subsubsection{Group C: cost-based algorithms} 

\textbf{C1} executes test cases in ascending order of their costs~\cite{cost-based}. Usually, the cost is the runtime of each test case (which can be estimated from the execution history). For example,  in Table~\ref{tab:example}, C1 will sort the test cases in the order \{t1, t2, t3, t4\} before Run 4 based on their increasing estimated cost from previous runs.

\subsubsection{Group D: test case description-based algorithms}

Group D algorithms utilize execution history information as dependent variables and test case description information as independent variables to build regression models and predict the probability that each test case will fail. The hypothesis behind this is that, if two test cases have similar descriptions, they tend to also have the same outcomes (fail/pass together). Text features are extracted from the test case description as described in \tion{method}. There are three ways to construct the dependent variables~\cite{59}:
\bi
\item
\textbf{D1:} Simple History (SH). Assign value 0 to the passed and skipped test cases and value 1 to the failed test cases in the immediately previous run. For example, the dependent variables of t1 to t4 of Run 4 in Table~\ref{tab:example} will be \{0,0,1,1\}. A regression model will be trained to learn that test cases are more likely to fail if ``page B'' is in their description. Therefore, D1 will sort the test cases in the order \{t3, t4, t1, t2\} before Run 4.
\item
\textbf{D2:} All History (AH). Assign the failure rate value (B2) as the dependent variable. This measure considers the entire
history rather than just the immediately previous run (SH). For example, the dependent variables of t1 to t4 of Run 4 in Table~\ref{tab:example} will be \{1/2, 2/3, 1/3, 2/3\}. A regression model will be trained to learn that test cases with ``Radio Button`` are more likely to fail than test cases with ``Check Box'' in their descriptions and test cases with ``page A'' are more likely to fail than test cases with ``page B'' in their descriptions. Therefore, D2 will sort the test cases in the order \{t2, t4, t1, t3\} before Run 4.
\item
\textbf{D3:} Weighted History (WH). Assign weighted averages (as
opposed to simple averages of AH measures) of SH values
in all previous runs placing more importance on recent
runs. For example, $\omega_k$ in \eqref{eq:ROCKET} from B4 can be applied as the weights and the dependent variables of t1 to t4 of Run 4 in Table~\ref{tab:example} will be \{0.2, 0.3, 0.7, 0.8\}. A regression model will be trained to learn that test cases with ``page B'' are more likely to fail than test cases with ``page A'' in their descriptions and test cases with ``Radio Button''are more likely to fail than test cases with ``Check Box'' in their descriptions . Therefore, D2 will sort the test cases in the order \{t4, t3, t2, t1\} before Run 4.
\ei

\subsubsection{Group E: feedback-based algorithms}

Group E algorithms utilize execution history information to find correlations between test cases, then dynamically order the test cases based on feedback information. For example, if t1 failed, then increase the priority of t3 because t1 and t3 are related. 
\bi
\item
\textbf{E1:} co-failure. The idea behind this is that, if test A and test B failed together in the past 75\% of the time, then if we observe a failure of test A
in the current run, we may prioritize the execution of test B since there would be a high probability that test B will fail as well~\cite{183}. Given a just finished test case result $t_{\text{finished}}$, for every unexecuted test case $t\in T_{\text{unexecuted}}$:
\begin{equation*}
\text{Priority}(t) = \text{Priority}(t) + (P(t=\text{fail}, t_{\text{finished}})-0.5)
\end{equation*}
where $P(t=fail, t_{\text{finished}})$ is the probability that $t$ fails given that $t_{\text{finished}}$ has passed/failed. For example, E1 initializes the priority of each test case as 0 for Run 4. Then it executes t1 first, and observes that t1 fails. The priority of t2, t3, and t4 will be updated to \{0.5, -0.5, -0.5\}. Therefore, t2 will be executed next. Given that t2 also fails, the priority of t3 and t4 will be updated to [-0.5, -1]. Then t3 will be executed and pass. The priority of t4 will be updated to -1 and finally t4 will be executed.
\item
\textbf{E2:} Flipping history. AFSAC~\cite{103} analyzes the execution history of the test suite and builts a correlation matrix of test cases. Two test cases are correlated when their results are changed to the opposite status by one commit in two consecutive test sessions (flipped together). The correlation matrix is composed of
values reflecting the accumulated number of flipped results. AFSAC~\cite{103} uses an improved ROCKET metric to find the first failure before reordering the test cases based on flipping history. For example, in Table~\ref{tab:example}, E2 initializes the order of Run 4 as \{t4, t3, t2, t1\}, then executes t4 and t4 fails. The priority of t1, t2, and t3 will be updated to \{1, 1, 1\} given that t1, t2, and t3 all flipped with t4 together once in the previous runs (runs 1 to 3). Therefore, there is no adjustment to the order and t3 is executed next. Since t3 passes, there is no update on the priorities and t2 will be executed. After t2 fails, the priority of t1 is updated to $\max\{1,1\} = 1$ since t1 also flipped with t2 together once. Finally t1 is executed.
\item
\textbf{E3:} Rule-based. REMAP~\cite{132} mines fail and pass rules over 90\% confidence with the Repeated Incremental Pruning to Produce Error Reduction (RIPPER) algorithm, and dynamically prioritizes the unexecuted test cases. If a test case fails, its corresponding fail rule test cases will be executed next; if a test case passes, its corresponding pass rule test cases will be pushed back to the end of the execution queue. The initial order of test cases is calculated by their failure rate (B2) and the number of rules associated. For example, in Table~\ref{tab:example}, with minimum support set to 2, one pass rule (if t2 fails then t3 will pass) and one fail rule (if t3 passes then t2 will fail) can be mined from Run 1 to Run 3. An initial order will be \{t2, t3, t4, t1\} given that t2 and t3 each has one rule associated and their failure rates are \{1/2, 2/3, 1/3, 2/3\}. After t2 fails, according to the pass rule, t3 is pushed back to the end of the execution order \{t4, t1, t3\}. Since there are no rules associated with t4 or t1, the order will not change after their execution.
\ei

\subsubsection{Group F: {\IT}}

Group F includes three variants of the proposed active learning based framework {\IT}. As shown in Table~\ref{tab:algorithms}, different variants of {\IT} utilize different information. 

Consider the example of Table~\ref{tab:example}, with $N_1=1$ and $N_2=2$. Here, {\IT} first randomly selects t2 to execute. After t2 fails, t1 is selected (randomly) as a presumptive non-relevant example, and a model is trained with t2 being positive (failure)  and t1 being negative (non-failure). Suppose the prediction probabilities of being positive for t1, t3, and t4 are \{0.1, 0.8, 0.6\}. Since $|R|<N_2$, uncertainty sampling is applied to select t4 for execution since the model is most uncertain about the prediction of t4 (closest to 0.5). After t4 fails, t3 and t1 are selected as presumptive non-relevant examples. After aggressive undersampling, a model is trained with t2 and t4 being positive, and with t1 and t3 being negative. Suppose the prediction probabilities of being positive for t1 and t3 are \{0.3, 0.2\}. Since $|R|\ge N_2$, certainty sampling is applied to select t1 for execution since t1 has a higher predicted probability for being positive than t3. 

Note that in the above example, F1, F2, and F3 will have different features; thus the predictions will be different, providing a different dynamic ordering of test cases.

\subsection{Dependent Variables}
\label{sec:metrics}

To answer the two research questions, we collected two performance metrics to evaluate the performance of each algorithm:
\bi
\item
\textbf{APFDc}: average percentage of failure detected with cost, as calculated in \eqref{eq:APFDc} with severities of each failure set to be the same ($f_i=1$). APFDc is applied to evaluate the failure detection rates and answer \textbf{RQ1}.
\item
\textbf{Overhead}: computation time of the algorithm / total runtime of all test cases. Overhead is collected to evaluate the extra computational cost of each TCP algorithm and answer \textbf{RQ2}.
\ei
To test the significance of any improvements, Scott-Knott analysis is applied to cluster and rank the two performance metrics of each algorithm from Run 6 to Run 54. It clusters algorithms with little difference in performance together and ranks each cluster with the median performances~\cite{scott1974cluster}. Nonparametric hypothesis tests are applied to handle the non-normal distribution. Specifically, Scott-Knott decided that, the two algorithms are not of little difference if both bootstrapping~\cite{efron1982jackknife}, and an effect size test (Cliff's delta)~\cite{cliff1993dominance}, agreed that the difference is statistically significant ($99\%$ confidence) and not a negligible effect (Cliff's Delta $\ge$ 0.147).

\begin{table*}[!b]
\caption{Results for the two metrics described in \tion{metrics}. Overhead is presented as percentages. Medians and IQRs show the 50th and (75-25)th percentile results for 49 simulations runs. Results are divided into ``ranks'' (shown in the left most columns). Results have the same rank if our statistical tests showed no significant different between them. Note that the result of E3 is not shown because it took too much time to mine the association rules from 2661 test cases. E3 ran for 30 hours and was still not finished. As a result, E3 has an overhead of more than 100\% and is considered not useful no matter how high an APFDc it can achieve.}
\label{tab:result}
\centering
\setlength\tabcolsep{5pt}
\subfloat[APFDc]
%{\footnotesize \begin{tabular}{l@{~~~}l@{~~~}r@{~~~}r@{~~~}c}
{\footnotesize \begin{tabular}{llllc}
\arrayrulecolor{lightgray}
\textbf{Rank} & \textbf{Treatment} & \textbf{Median} & \textbf{IQR} & \\\hline
  1 &           E2 &    0.49  &  0.08 & \quart{0}{12}{4}{84} \\
  1 &           A1 &    0.50  &  0.01 & \quart{4}{2}{6}{84} \\
  1 &           C1 &    0.50  &  0.03 & \quart{1}{5}{6}{84} \\
  1 &           E1 &    0.52  &  0.05 & \quart{4}{8}{9}{84} \\
\hline  2 &           D1 &    0.60  &  0.15 & \quart{12}{24}{21}{84} \\
  2 &           D3 &    0.61  &  0.11 & \quart{17}{17}{23}{84} \\
  2 &           F1 &    0.61  &  0.07 & \quart{17}{11}{23}{84} \\
  2 &           D2 &    0.62  &  0.10 & \quart{17}{15}{25}{84} \\
  2 &           B1 &    0.63  &  0.19 & \quart{15}{30}{26}{84} \\
\hline  3 &           F2 &    0.66  &  0.18 & \quart{17}{28}{31}{84} \\
  3 &           B5 &    0.67  &  0.11 & \quart{23}{17}{32}{84} \\
  3 &           B2 &    0.67  &  0.11 & \quart{23}{17}{32}{84} \\
  3 &           B4 &    0.67  &  0.17 & \quart{21}{27}{32}{84} \\
  3 &           B3 &    0.67  &  0.18 & \quart{20}{28}{32}{84} \\
\hline  4 &           F3 &    0.73  &  0.13 & \quart{26}{21}{42}{84} \\
\hline  5 &           A2 &    0.95  &  0.09 & \quart{65}{14}{76}{84} \\
\hline \end{tabular}}
\quad\quad\quad\quad
\setlength\tabcolsep{5pt}
\subfloat[Overhead]
%{\footnotesize \begin{tabular}{l@{~~~}l@{~~~}r@{~~~}r@{~~~}c}
{\footnotesize \begin{tabular}{llllc}
\arrayrulecolor{lightgray}
\textbf{Rank} & \textbf{Treatment} & \textbf{Median} & \textbf{IQR} & \\\hline
  1 &           A2 &    0.02  &  0.01 & \quart{0}{0}{0}{12} \\
\hline  2 &           A1 &    0.2  &  0.1 & \quart{1}{1}{2}{12} \\
  2 &           C1 &    0.21  &  0.07 & \quart{1}{1}{2}{12} \\
  2 &           B1 &    0.21  &  0.11 & \quart{1}{1}{2}{12} \\
  2 &           D1 &    0.22  &  0.08 & \quart{1}{1}{2}{12} \\
  2 &           B4 &    0.22  &  0.11 & \quart{1}{1}{2}{12} \\
  2 &           B3 &    0.22  &  0.10 & \quart{1}{1}{2}{12} \\
  2 &           B2 &    0.22  &  0.08 & \quart{1}{1}{2}{12} \\
  2 &           D2 &    0.22  &  0.10 & \quart{1}{1}{2}{12} \\
  2 &           B5 &    0.22  &  0.12 & \quart{1}{2}{2}{12} \\
  2 &           D3 &    0.22  &  0.11 & \quart{1}{1}{2}{12} \\
\hline  3 &           F2 &    0.23  &  0.07 & \quart{2}{0}{2}{12} \\
  3 &           E2 &    0.24  &  0.07 & \quart{2}{1}{2}{12} \\
\hline  4 &           F3 &    0.33  &  0.11 & \quart{3}{1}{3}{12} \\
  4 &           F1 &    0.34  &  0.10 & \quart{3}{2}{4}{12} \\
\hline  5 &           E1 &    5.79  &  2.52 & \quart{48}{31}{72}{12} \\
\hline \end{tabular}}

\end{table*}

\subsection{Threats to validity}
\label{sec:validity}

This section discusses validity threats~\cite{feldt2010validity} to the above design. Any conclusions made from this work must be considered with the following issues in mind:

\textbf{External validity} concerns how well the conclusion can be applied outside. All the conclusions in this study are drawn from 54 automated UI testing runs from LexisNexis. When applied to other case studies, the following concerns might arise: 1) {\IT} might not be applicable if one of the three types of information is not available; 2) other TCP algorithms might perform better than {\IT} if more information is available, e.g. coverage-based algorithms.

\textbf{Internal validity }focuses on how sure we can be that the treatment caused the outcome. To enhance   internal validity, we heavily constrained our simulations to the same dataset with the same workflow for the TCP algorithms we evaluate. However, some of the algorithms in Table~\ref{tab:algorithms} are not exact implementations or reproductions of the original algorithms. Rather, they are only partially implemented to fit the workflow and information availability of our data. This may be a threat to internal validity. Another threat to internal validity comes from flaky tests. Apart from removing all the ``time out'' failures, we did not identify other types of flaky tests. These flaky tests could have affected the model learned as well as the evaluations. We plan to explore ways to tackle flaky tests problem in our future work.

\textbf{Construct validity }focuses on the relation between the theory behind the experiment and the observation. We use default parameters for the 17 algorithms in our simulations. For some of the algorithms, e.g. B3 and D1, different parameter settings can make large differences in failure detection rates. Therefore, it is possible that our observation does not hold when different parameters are applied for each algorithm. We would consider hyper-parameter tuning as future work to alleviate this concern.

\textbf{Conclusion validity} focuses on the significance of the treatment. To enhance   conclusion validity, we applied Scott-Knott tests to the collective results of 49 runs to see whether one TCP algorithm was significantly better than another on  certain performance metrics.

\subsection{Results}

Table~\ref{tab:result} shows the results of the 17 algorithms presented in Table~\ref{tab:algorithms}. For each algorithm, APFDc and overhead from 49 runs are recorded and their medians and iqrs are presented. Two algorithms perform significantly different in terms of one metric if they are in different ranks of the Scott-Knott analysis. APFDc evaluates failure detection rates, with higher numbers being better, while overhead represents the extra cost of running the algorithms, with lower numbers being better. 

\begin{figure}
    \centering
    \includegraphics[width=\linewidth]{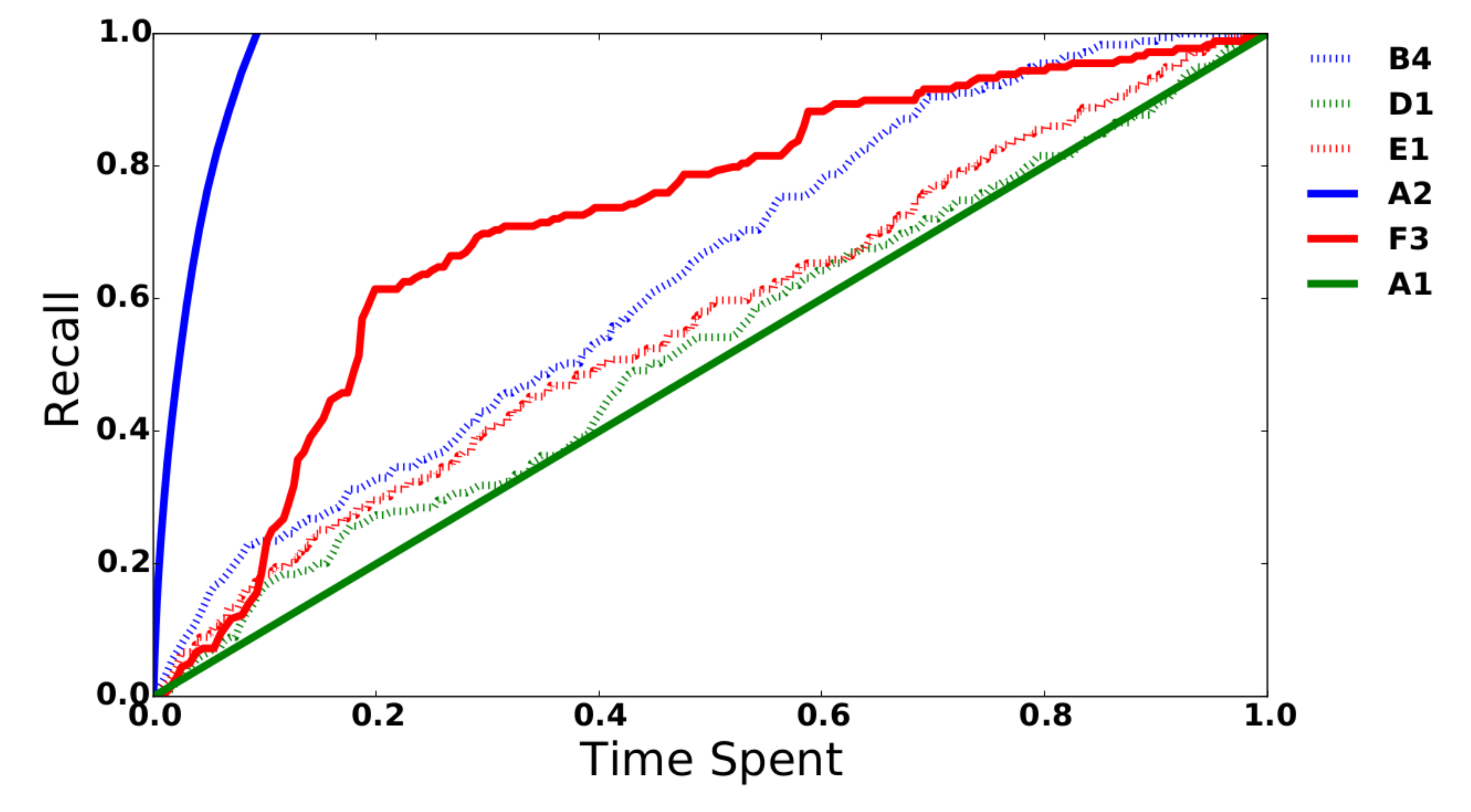}
    \caption{Failure detection curve of an example run at 02:02:34, October 24th, 2018. Here recall is the number of failures detected divided by the total number of failures in the test suite. An algorithm is considered better than another if it achieves higher recall with less time spent.}
    \label{fig:curve}
\end{figure}

First, we analyze the APFDc results from Table~\ref{tab:result} for \textbf{RQ1}:
\bi
\item
F3, which is {\IT} with the hybrid feature, performed the best in terms of APFDc. Its failure detection rates are significantly higher than those of the other algorithms (9\% higher than a Rank 3 algorithm like B4). 
\item
Simple history-based algorithms (B2, B3, B4, B5) are the second best algorithms in terms of APFDc. The more complex algorithms from Groups D or E that utilized more information performed worse than the simple history-based algorithms from Group B.
\ei
Even though 9\% higher APFDc does not seem to be a large improvement, it actually increases the failure detection rates substantially. As an example, Figure~\ref{fig:curve} shows the failure detection curves for the algorithms of each rank. Here we can see that {\IT} (F3) finds many more failures earlier than the other TCP algorithms, e.g., when 20\% of the time is spent, F3 finds 60\% of the failures while B4 finds only 30\%. By comparing the A2 and F3 results from Table~\ref{tab:result}, we can offer a precise measure for the success of our new proposed method:

\begin{RQ}{RQ1: can {\IT} achieve significantly higher failure detection rates than other TCP algorithms when prioritizing automated UI tests from LexisNexis?}
As seen in the {\bf Rank} column of Table~\ref{tab:result}, F3 ({\IT} with the hybrid feature) performed significantly better than other
TCP methods.   Further, {\IT} performed within \mbox{73/95 = 75\%} of the  optimal (A2) in median.
\end{RQ}

Second, we analyze the overhead results from Table~\ref{tab:result} for \textbf{RQ2}:
\bi
\item
Even though F3 has higher overhead (50\% higher than simple history-based algorithms from Group B), its overhead (35 seconds in median, which is only 0.33\% of the total runtime of the test suite) is negligible compared to the cost it can potentially save.
\item
Algorithms utilizing feedback information (Groups E and F) tend to have higher overhead since the orders of test cases are dynamically adjusted based on the execution results. \ei

\begin{RQ}{RQ2: what is the computational overhead of {\IT}?}
The computational overhead of {\IT} is $0.33\%$ of the total runtime of the test suite, which is negligible compared to the cost it can potentially save  by achieving higher failure detection rates.
\end{RQ}

We explain the superior performance of {\IT} as follows:
\begin{enumerate}
\item
{\IT} has an advanced framework adapted from the solutions of the total recall problem which utilizes different data balancing techniques and query strategies.
\item
{\IT} utilizes more information (test history, test case description, and current feedback) than other algorithms.
\end{enumerate}
That said, there are some limitations to the current results. Specifically, compared to the optimal result, the APFDc of F3 is still lower than that of A2. That is, there  is still room to improve this result.

\section{organizational impacts}
\label{sec:impact}

With higher failure detection rates provided by {\IT},  LexisNexis teams can  build more versions of tested software within one eight hour shift.  This will have three major effects:
 \bi
 \item
{\em  Higher responsiveness to customer feedback.} The faster LexisNexis  can build
 and test systems, the faster their customers can see  new features (or fixes to old features).
 \item
{\em  Greater programmer adaptability.}
The faster programmers can get feedback on their systems, the better they can become in software  adaption and extension. 
\item {\em Increased recruitment and retention of programmers.}  
Here at NC State, many of our graduates work in the large and local 
information technology industry. Based on feedback from that population,
we can assert that organizations
that ship products faster can also recruit and retain
more software developers. 
\ei

\section{Conclusion and future work}
\label{sec:conclusion}

Automated UI testing is an important component of the continuous integration process of software product development. Faster automated UI testing leads to shorter development cycles and better quality. This paper studies the specific problem of how to prioritize automated UI test cases. By analyzing what information is available when prioritizing automated UI test cases, 12 state of the art TCP algorithms are found to be applicable to automated UI test case prioritization. In addition, prioritizing test cases for failure detection rates can be generalized as a total recall problem. A new TCP algorithm is proposed by adapting the active learning based framework from the total recall problem to the TCP problem. Utilizing the information of execution history, test case description, and feedback, {\IT} (F3) outperformed all other TCP algorithms in terms of failure detection rates (APFDc) on a dataset of 54 consecutive runs of real world automated UI testing in LexisNexis. Lessons learned from this work are summarized as follows:
\bi
\item
Automated UI testing is widely applied in industry and consumes a large portion of time and resources. Optimizing the execution of automated UI testing is a research problem that deserves much more   attention.
\item
Black box test case prioritization (given no information on source code) for failure detection rates is under-explored in the research arena.
This is strange since, at least in our experience,
this kind of testing  is more often than not the only  practical approach for large organizations like LexisNexis.
\item
The test case prioritization for failure detection rates problem can be generalized as a total recall problem. Techniques from the total recall problem can be applied to improve performance on the TCP problem. This also suggests that future improvements in TCP can in return be applied to help solve other total recall problems like citation screening, vulnerability inspection, and static warning identification.
\ei
Given the validity threats and limitations of current work, our future work includes:
\begin{enumerate}
\item
Hyper-parameter tuning for different TCP algorithms including the proposed active learning based framework to see if better configurations can be found.
\item
Identify flaky test cases and avoid their impact on the learned active learning model.
\item
It is tedious and time-consuming to classify each test failure to a specific fault caused by some piece of the source code. The study of fault localization might help address this problem.
\item
Apply {\IT} to other TCP problems.
\item
Follow up with LexisNexis on how to integrate the proposed framework with their current automated UI testing system to obtain higher failure detection rates.
\end{enumerate}

\section*{Acknowledgement}
This work was partially funded by a gift from LexisNexis
managed by Phillpe Poignant and an NSF CCF grant \#1703487.

% Bibliography
\balance
\bibliographystyle{ACM-Reference-Format}
\bibliography{mybib,literature}

\end{document}